# Laser ablation of $Fe_2B$ target enriched in $^{10}B$ content for boron neutron capture therapy

Ekaterina V. Barmina[1], Irina N. Zavestovskaya[2,3], Anna I. Kasatova[5], Dmitry S. Petrunya[2,3], Oleg V. Uvarov[1], Vladimir V. Saraykin[4], Margarita I. Zhilnikova[1], Valery V. Voronov[1], Georgy A. Shafeev[1,3], Sergey Yu. Taskaev[5,6]

[1]Prokhorov General Physics Institute of the Russian Academy of Sciences, 38, Vavilov street, 119991, Moscow, Russian Federation

[2]Lebedev Physical Institute of the Russian Academy of Sciences, 53, Leninsky Pr., Moscow 119991, Russian Federation

[3]National Research Nuclear University MEPhI (Moscow Engineering Physics Institute), 31, Kashirskoye highway, 115409, Moscow, Russian Federation

[4]National Research Center «Kurchatov Institute», Physical Problems Research Institute, 5, Georgievsky prospekt, 124460, Moscow, Zelenograd, Russian Federation

[5]Budker Institute of Nuclear Physics of Siberian Branch Russian Academy of Sciences, 11, Acad. Lavrentieva Pr., Novosibirsk, 630090 Russian Federation

[6]Novosibirsk State University, 1, Pirogov street, Novosibirsk, 630090 Russian Federation

**Abstract**

The technique of laser ablation in liquids is applied to produce of Boron-containing nanoparticles from ablation of a $Fe_2B$ bulk target enriched in $^{10}B$ isotope. Laser ablation of the target in liquid isopropanol results in partial disproportionation to free Fe and Boron while nanoparticles of $Fe_2B$ are also presented. Nanoparticles are magnetic and can be collected using a permanent magnet. Average size of nanoparticles is of 15 nm. The content of $^{10}B$ in generated nanoparticles amounts to 76,9 %. Nanoparticles are biocompatible and can be used in Boron Neutron Capture Therapy.

The Neutron Capture Therapy (NCT) is an actively developing direction of binary technologies of radiation therapy [1]. The NCT method is used for the treatment of inoperable and radioresistant cancers, and in cases where other methods of treatment are ineffective. It is based on the preliminary saturation of cancer cells with elements having a high neutron capture cross-section, and subsequent irradiation of them with low-energy (epithermal) neutrons. The most effective is Boron-NCT (BNCT) when to efficiently capture thermal neutrons is used one of non-radiative boron isotope, $^{10}$B [1, 2]. Then the isotope $^{11}$B is formed that almost instantly decays to $^7$Li nuclei and an α-particle with a high linear energy transfer. Both the α-particles and the lithium nuclei produce closely spaced ionizations in the immediate vicinity of the reaction. The mean free path of α-particles and lithium nuclei in biological tissues is of several micrometers, which is comparable with the size of a target biological cell. As a result, the absorption of energy is localized and leads to the death of cancer cells while the cells of healthy tissues are not affected by neutrons. Currently, several boron-containing chemical compounds are used for BNCT procedures. The main disadvantage of these compounds is relatively low boron content. It is desirable to increase the boron content in the used compounds.

To increase efficiency of BNCT some groups use boron-containing nanoparticles that have higher value of B atoms compared to boronophenylalanine (BPA) and sodium borocaptate (BSH) [3-8]. To produced such functional nanoparticles chemical methods are used. It leads to formation of NPs with different kinds of additional ions and chemical impurities that makes them impossible to apply directly for biological cells.

Laser ablation is one of the method for formation of pure nanoparticles what is especially important for biomedical applications [9-12]. In this case irradiation of target is performed usually in water that guarantees generation of NPs without any outsider components.

The best candidates for NCT are not only NPs of Boron itself but also its binary compounds, for example, $Fe_2B$. Fe-B materials doped with rare earth elements, such as Nd, are the strongest permanent magnets. One may expect that NPs of such materials obtained with the help of laser ablation in liquids will retain magnetic properties of the starting material. This would allow the use of external magnets to localize the nanoparticles in the place where they are needed.

Laser ablation of $Fe_2B$ target with natural composition in Boron isotopes in liquid acetone has been successfully realized [13]. These NPs are magnetic and can be collected using a permanent magnet. However, even with NPs of $Fe_2B$ the Boron content is about 10%. The $^{10}$B content in natural Boron is of order of 20%, so the content of $^{10}$B active in BNCT process is of 2% by mass of $Fe_2B$ NPs.

In this work we present the results on laser ablation of a $Fe_2B$ target enriched in $^{10}$B content with final aim using these NPs in BNCT.

The radiation of an Ytterbium fiber laser with a wavelength of 1060-1070 nm was focused by an objective (F = 20.4 cm) onto the surface of the $Fe_2B$. The laser beam moved along the target surface at the speed of 100-500 mm/s by galvano system mirrors. Pulse duration was of 200 ns and laser fluence on the target was around 6 J/cm$^2$.

Two liquids were tested as the medium for laser ablation, either water $H_2O$ or isopropanol $C_3H_8O$. The target had a shape of a massive cylinder made of $Fe_2B$ of 35 mm in diameter preliminary enriched in $^{10}B$ content using a photonuclear reaction with gamma photons. Polymeric walls of the cell were fixed on the target itself and a soda lime glass transparent for laser radiation was fixed on walls at the height of 5 mm above the surface of the cylinder. The cell was included into a flow cell equipped with peristaltic pump.

The NPs morphology was analyzed with a Carl Zeiss 200FE transmission electron microscope (TEM). Diffraction patterns (XRD) of NPs were recorded using X-ray diffractometer Bruker D8 Discover A25 Da Vinci Design, $CuK_\alpha$ radiation with $\lambda=1.5418$Å. Isotopic composition of the generated NPs was characterized by mass-spectrometer of secondary ions IMS-4f, CAMECA.

Laser ablation is accompanied by the liquid breakdown above the target surface. Laser ablation of $Fe_2B$ target in $H_2O$ results in its decomposition to presumably iron hydroxides and formation of some other NPs. These NPs are rusty in appearance, which is probably due to the interaction of Fe with $H_2O$. The NPs are not magnetic, which means that they contain neither Fe nor $Fe_2B$. Extinction spectra are shown in Fig. 1.

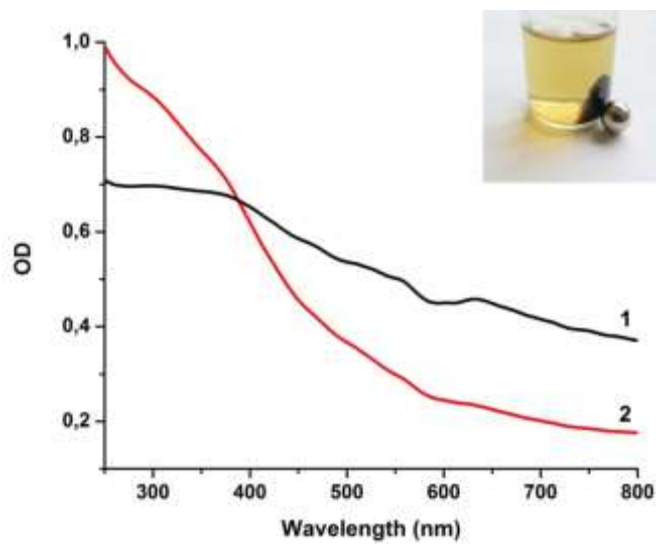

Fig. 1. Extinction spectra of colloidal solutions of NPs ablated in isopropanol and then replaced by $H_2O$ (1) and ablated in $H_2O$ (2). In the inset: collection of NPs with a magnet.

The NPs generated by laser ablation in isopropanol are black in appearance. They are magnetic and can be collected using a permanent magnet. Since biological tests require an aqueous medium for $Fe_2B$ NPs, isopropanol was substituted by $H_2O$ in several cycles. The generated NPs were evaporated on a Si substrate for further mass-spectrometric analysis.

X-ray diffractogram of the initial bulk $Fe_2B$ target is presented in Fig. 2,a. It coincides with the diffraction pattern for $Fe_2B$ from database Powder Diffraction File-2, version 2011. X-ray diffractogram of the NPs produced by laser ablation of this target in isopropanol are shown in Fig.2, b.

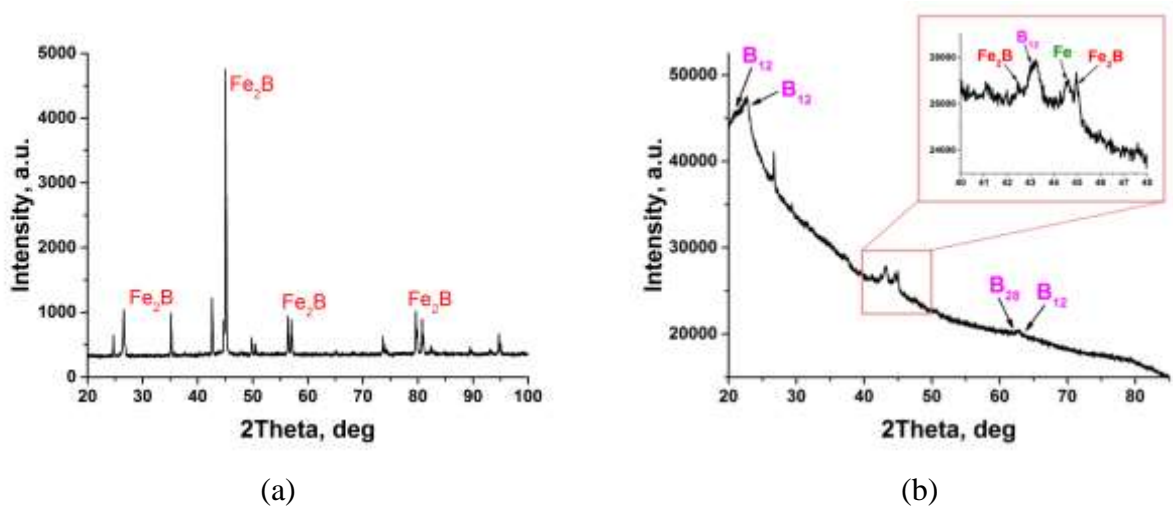

(a) (b)

Fig. 2. X-ray diffractograms of the initial Fe$_2$B target (a) and NPs generated by laser ablation of this target in liquid isopropanol (b).

High background in Fig. 2, b is due to X-ray luminescence induced by CuK$_\alpha$ radiation in Fe. According to XRD data, Fe$_2$B compound is partly disproportionated upon laser ablation in isopropanol. Indeed, except for Fe$_2$B NPs there are also NPs of Fe and stable boron molecules B$_{12}$ and B$_{28}$ made of 12 and 28 atoms of Boron, respectively.

TEM view of nanoparticles generated by laser ablation of Fe$_2$B target in isopropanol is shown in Fig. 3.

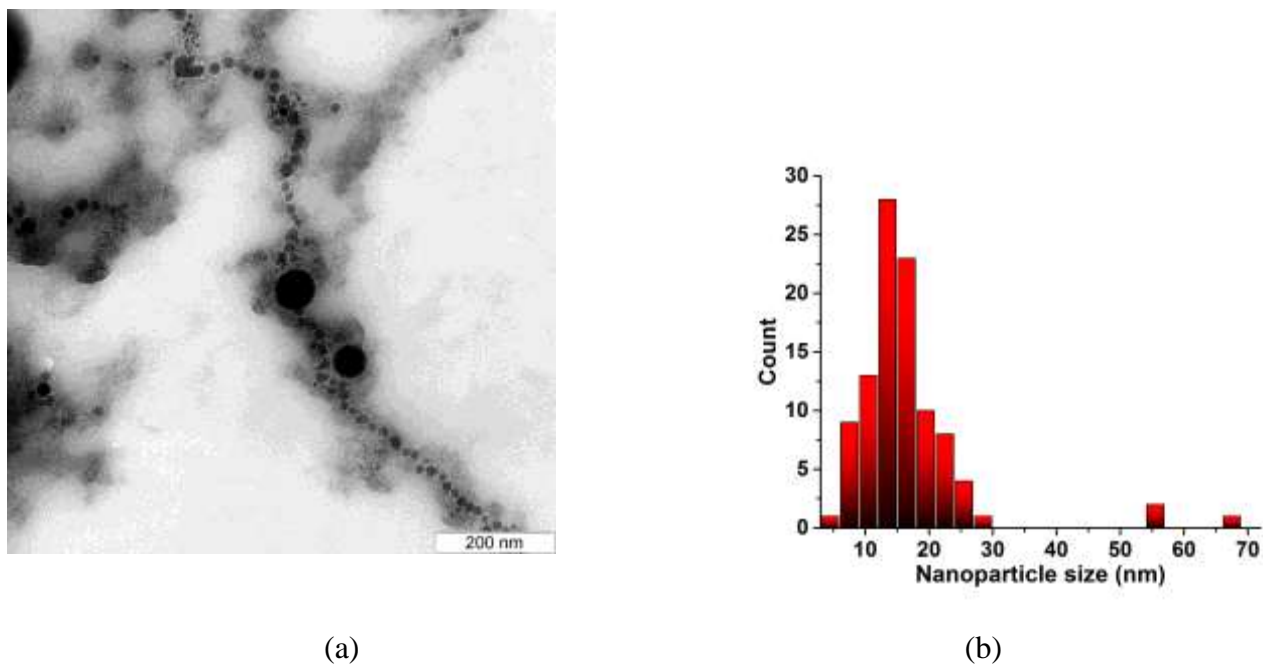

(a) (b)

Fig. 3. TEM view of NPs produced by laser ablation of Fe$_2$B target in isopropanol. Scale bar denotes 200 nm (a). Histogram of size distribution of NPs in a, (b).

One can see that NPs are aligned one after another which confirms their magnetism. The aligned NPs are embedded into amorphous halo that presumably consists of the products of $Fe_2B$ decomposition. Amorphous carbon may also be presented since it is formed under laser decomposition of isopropanol. This carbon should not have diffraction peaks in XRD.

View in scattered electrons (STEM) of NPs prepared by laser ablation of $Fe_2B$ target in isopropanol is shown in Fig. 4.

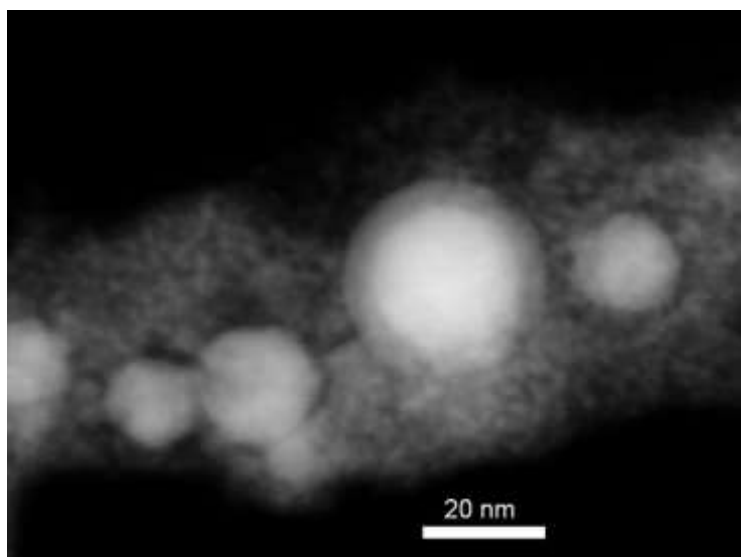

Fig. 4. STEM view of NPs obtained by laser ablation of $Fe_2B$ in isopropanol. Scale bar denotes 20 nm.

One can see dense core and less dense shell around it. Amorphous halo is also made of less dense materials than the core.

Isotopic composition of NPs obtained by laser ablation of $Fe_2B$ target in isopropanol was compared with that of industrial Boron powder fragmented by laser radiation in $H_2O$. The results are shown in Table 1.

Table 1

| Boron isotope content, % | $^{10}B$ | $^{11}B$ |
|---|---|---|
| Industrial B powder | 19.2±0.1 | 81.8±0.1 |
| Ablated $Fe_2B$ nanoparticles | 76.9±0.1 | 23.1±0.1 |

One can see that the content of $^{10}$B in NPs generated by laser ablation of Fe$_2$B target enriched in this isotope is almost 4 times higher $^{10}$B content in industrial Boron powder. This is advantageous for required for BNCT amount of NPs.

The particles were successfully tested for cytotoxicity on two types of cell cultures U87 (human glioblastoma) and SW-620 (human colorectal adenocarcinoma).

Thus, NPs enriched in $^{10}$B content have been successfully realized by laser ablation of a bulk Fe$_2$B target in isopropanol. It was found that NPs contain apart from Fe$_2$B NPs also NPs of Fe and molecular B$_{12}$ and B$_{28}$ structures. Average size of NPs is about 15 nm, and some of them have core-shell structure. NPs are magnetic and can be collected and directed using a permanent magnet. The content of $^{10}$B in laser-produced nanoparticles is as high as 76.9%, which exceeds the $^{10}$B content in natural Boron almost by factor 4. The generated nanoparticles are biocompatible and can be used in BNCT.

**Acknowledgements**

This work was supported by the grants of the Russian Foundation for Basic Research (20-32-70112-Stabil'nost') and RF President's Grant MD-3790.2021.1.2. This work was partly performed within the framework of National Research Nuclear University MEPhI (Moscow Engineering Physics Institute) Academic Excellence Project (Contract No. 02.a 03.21.0005). We are also grateful to Common Use Center of GPI for X-ray data and TEM images of NPs.